\documentclass[12pt]{article}
\usepackage{graphicx}

\begin{document}

\title{Algorithm for planar 4-Body Problem central configurations with given masses}

\author{E. Pi\~na \\
Department of Physics \\ Universidad Aut\'onoma Metropolitana - Iztapalapa, \\
P. O. Box 55 534. Mexico, D. F., 09340 Mexico \\
e-mail: pge@xanum.uam.mx}

\date{ }
\maketitle

\abstract{
An algorithm to compute the six distances between particles of a planar Four-Body central configuration is presented according to the following schema. An orthocentric tetrahedron is computed as a function of given masses. Each mass is placed at the corresponding vertex of the tetrahedron. The center of mass (and orthocenter) of the tetrahedron is at the origin of coordinates. The tetrahedron is orientated in a particular position function of the masses: with one of the particles placed on axis 3. The tetrahedron is rotated by two angles (to be tuned variables) around the center of mass until a direction orthogonal to the plane of configuration coincides with axis 3. The four coordinates of the vertices of the tetrahedron along this direction are identified with the weighted directed areas of the central configuration. The central configuration corresponding to these weighted directed areas is computed giving rise to four masses and corresponding distances. The given masses are compared with the computed ones. The two angles of the rotation are tuned until the given masses coincide with the computed. The corresponding distances of this last computation determine the central configuration. The case with two equal masses also is considered.
}
\


\


\newpage

\section{Introduction}

The coordinate system used in this paper has been presented in reference  \cite{pi}. The algorithm to compute planar central configurations starting from the weighted areas was published with several applications in \cite{pl}.

To begin I make a review of the new four body coordinates that mimics an important portion of \cite{pi}, expanded at important points and reduced in several comments.

The masses of the four bodies $m_1$, $m_2$, $m_3$ and $m_4$ are positive, generally different, but the values could be repeated.

We transform from the inertial referential, to the frame of principal axes of inertia by means of a three dimensional rotation $\bf G$ parameterized by three independent coordinates.

In addition to this rotation three more coordinates are introduced, as scale factors $R_1$, $R_2$, $R_3$, which are three directed distances closely related to the three principal inertia moments through
\begin{equation}
I_1 = \mu (R_2^2 + R_3^2)\, , \quad I_2 = \mu (R_3^2 + R_1^2) \, , \quad \mbox{and} \quad I_3 = \mu (R_1^2 + R_2^2) \, ,
\end{equation}
where $\mu$ is the mass
\begin{equation}
\mu = \sqrt[3]{\frac{m_1 \, m_2\, m_3\, m_4}{m_1 + m_2 + m_3 + m_4}} \, .
\end{equation}

With the first rotation and the change of scale the resulting four-body configuration has a moment of inertia tensor with the three principal moments of inertia equal. The second rotation $\bf G'$ does not change this property.

The cartesian coordinates of the four particles, with the center of gravity at the origin, written in terms of the new coordinates are
\begin{equation}
\left(\begin{array}{cccc}
x_1 & x_2 & x_3 & x_4 \\
y_1 & y_2 & y_3 & y_4 \\
z_1 & z_2 & z_3 & z_4
\end{array} \right) = {\bf G} \left(\begin{array}{ccc}
R_1 & 0 & 0 \\
0 & R_2 & 0 \\
0 & 0 & R_3
\end{array} \right) {\bf G'}^{\rm T}
\left(\begin{array}{cccc}
a_1 & a_2 & a_3 & a_4 \\
b_1 & b_2 & b_3 & b_4 \\
c_1 & c_2 & c_3 & c_4
\end{array} \right)\, ,
\end{equation}
where $\bf G$ and $\bf G'$ are two rotation matrices, each one a function of three independent coordinates such as the Euler angles, and where the column elements of the constant matrix
\begin{equation}
{\bf E} = \left(\begin{array}{cccc}
a_1 & a_2 & a_3 & a_4 \\
b_1 & b_2 & b_3 & b_4 \\
c_1 & c_2 & c_3 & c_4
\end{array} \right)\, ,
\end{equation}
are the coordinates of the four vertices of a rigid orthocentric tetrahedron, with the center of mass at the origin of coordinates, namely:
\begin{equation} 
\begin{array}{c}
a_1 m_1 + a_2 m_2 + a_3 m_3 + a_4 m_4 = 0 \, , \\
b_1 m_1 + b_2 m_2 + b_3 m_3 + b_4 m_4 = 0 \, , \\
c_1 m_1 + c_2 m_2 + c_3 m_3 + c_4 m_4 = 0 \, .
\end{array}\, . \label{base}
\end{equation}

We introduce the following notation for the matrix
\begin{equation}
{\bf M} = \left( \begin{array}{cccc}
m_1 & 0 & 0 & 0 \\
0 & m_2 & 0 & 0 \\
0 & 0 & m_3 & 0 \\
0 & 0 & 0 & m_4
\end{array} \right) \, .
\end{equation}

An equivalent condition in order to have three equal inertia moments for the rigid tetrahedron is expressed as
\begin{equation} 
{\bf E\, M\, E}^{\rm T} = \mu \left( \begin{array}{ccc}
1 & 0 & 0 \\
0 & 1 & 0 \\
0 & 0 & 1
\end{array} \right) \label{ort}
\end{equation}

An orthocentric tetrahedron has the property that the perpendicular lines to the faces trough the four vertices intersect at the same point. Orthocentric tetrahedra were considered by Lagrange in 1773 \cite{la}. Other old references on orthocentric tetrahedra are found in a paper by Court \cite{co}, he calls them orthocentric and orthogonal to these tetrahedra. Placing the four masses at the corresponding vertices, the intersection point is actually the center of mass of the four masses, and the moment of inertia tensor of the four particles has the same principal value in any direction. Equation (\ref{ort}) imply that the inertia tensor of the rigid tetrahedron is proportional by a factor $2 \mu$ to the unit matrix.

To show these properties we consider for a while a fourth coordinate for the four vertex of the tetrahedron
\begin{equation}
(d_1, d_2, d_3, d_4) = \sqrt{\frac{\mu}{m}} (1, 1, 1, 1)\, , \label{cuarto}
\end{equation}
where we use the notation $m = m_1 + m_2 + m_3 + m_4$ for the total mass of the system. Then using properties (\ref{base}), (\ref{ort}), and (\ref{cuarto}) we write them in terms of $r = \sqrt{\frac{\mu}{m}}$ in the form
\begin{equation}
\frac{1}{\mu} \left( \begin{array}{cccc}
a_1 & a_2 & a_3 & a_4 \\
b_1 & b_2 & b_3 & b_4 \\
c_1 & c_2 & c_3 & c_4 \\
r & r & r & r
\end {array} \right) {\bf M} \left( \begin{array}{cccc}
a_1 & b_1 & c_1 & r \\
a_2 & b_2 & c_2 & r \\
a_3 & b_3 & c_3 & r \\
a_4 & b_4 & c_4 & r
\end {array} \right) = \left( \begin{array}{cccc}
1 & 0 & 0 & 0 \\
0 & 1 & 0 & 0 \\
0 & 0 & 1 & 0 \\
0 & 0 & 0 & 1
\end {array} \right)\, .
\end{equation}
Since the inverse matrix from the left is equal to the inverse from the right, this equation transforms into
\begin{equation}
\left( \begin{array}{cccc}
a_1 & b_1 & c_1 & r \\
a_2 & b_2 & c_2 & r \\
a_3 & b_3 & c_3 & r \\
a_4 & b_4 & c_4 & r
\end {array} \right) \left( \begin{array}{cccc}
a_1 & a_2 & a_3 & a_4 \\
b_1 & b_2 & b_3 & b_4 \\
c_1 & c_2 & c_3 & c_4 \\
r & r & r & r
\end {array} \right) = \left( \begin{array}{cccc}
\frac{\mu}{m_1} & 0 & 0 & 0 \\
0 & \frac{\mu}{m_2} & 0 & 0 \\
0 & 0 & \frac{\mu}{m_3} & 0 \\
0 & 0 & 0 & \frac{\mu}{m_4}
\end {array} \right)\, .
\end{equation}
This matrix equation is equal to its transposed; therefore it just has ten independent equations. Four of them are
\begin{equation}
a_j^2 + b_j^2 + c_j^2 = \mu \left( \frac{1}{m_j} - \frac{1}{m}\right)\, \quad (j=1, 2, 3, 4)\, , \label{20}
\end{equation}
for the square of the distance from the orthocenter to the vertex $j$.
The other six are
\begin{equation}
a_i a_j + b_i b_j + c_i c_j = - \frac{\mu}{m} \quad \mbox{ ($i \neq j$)}\, , \label{21}
\end{equation}
the internal product of two directed vertices is the same for the six different possible choices of the pair $i, j$.
From these basic equations it is easy to show that the position vector of one vertex is orthogonal to the three vectors between two vertices of the oposite face (to the position vector of the vertex.)
\begin{equation}
a_i (a_j - a_k) + b_i (b_j - b_k) + c_i (c_j - c_k) = 0 \quad \mbox{ ($i,j,k$ different.) }
\end{equation}
In addition, the square of the distance between two vertices is given by
\begin{equation}
(a_i - a_j)^2 + (b_i - b_j)^2 + (c_i - c_j)^2 = \mu \left( \frac{1}{m_i} + \frac{1}{m_j}\right)\, . \label{lados}
\end{equation}
This last is the condition to have a moment of inertia tensor with the same three principal moments of inertia. The six edges of the tetrahedron should be equal (for an arbitrary $\mu$) to the square root of the right-hand side of this equation. The volume of this tetrahedron is equal to 1/6 if $\mu$ is selected as above.

There are other remarkable geometrical properties of an orthocentric tetrahedron. The center of mass of each face is at the orthocenter where the three altitudes of the face intersect. This point is on the same straight line between the opposite vertex and the center of mass. In addition to the orthogonality of the three sets of two opposite edges of the tetrahedron, the two orthogonal edges are also orthogonal to the line joining the center of mass of the two edges.

There are important coordinate systems to fix the origin for measuring the $\bf G'$ rotation; from these I prefer to choose in this paper particle with mass $m_1$ along coordinate axis 3, the other three in a parallel plane to the coordinate plane of 1 and 2, which does not include the  first particle; the particle with mass $m_2$ on an orthogonal coordinate plane that includes the first particle and the center of mass, and the other two particles on a line that is parallel to the coordinate axis 1 and perpendicular to the coordinate plane of the first two particles. Particle 1 has the coordinates
\begin{equation}
(a_1, b_1, c_1) = \left(0, 0, \sqrt{\frac{\mu (m - m_1)}{m_1 m}}\right)\, .
\end{equation}
Particle 2 has the coordinates
\begin{equation}
(a_2, b_2, c_2) = \left(0, \sqrt{\frac{\mu (m_3+m_4)}{m_2 (m - m_1)}}, - \sqrt{\frac{\mu m_1}{(m - m_1) m}}\right)\, .
\end{equation}
Particle 3 has the coordinates
$$
(a_3, b_3, c_3) =
$$
\begin{equation}
\left(\sqrt{\frac{\mu m_4}{m_3 (m_3+m_4)}}, - \sqrt{\frac{\mu m_2}{(m_3+m_4) (m - m_1)}}, - \sqrt{\frac{\mu m_1}{(m - m_1) m}}\right)\, .
\end{equation}
Particle 4 has the coordinates
$$
(a_4, b_4, c_4) =
$$
\begin{equation}
\left(- \sqrt{\frac{\mu m_3}{m_4 (m_3+m_4)}}, - \sqrt{\frac{\mu m_2}{(m_3+m_4) (m - m_1)}}, - \sqrt{\frac{\mu m_1}{(m - m_1) m}}\right)\, .
\end{equation}

This rigid tetrahedron is the generalization of the rigid triangle of the Three-Body problem with the center of mass at the orthocenter discussed previously in \cite{pb}. The same triangle was used with different purposes by C. Sim\'o \cite{si}.

I assume for simplicity that the potential energy is given by the Newton potential (the gravitational constant is equal to 1)
\begin{equation}
V = - \sum_{i<j}^3\frac{m_i m_j}{r_{ij}}\, ,
\end{equation}
although our results may be generalized for any potential with a given power law of the relative distances between particles $r_{ij}$. It follows the relation between the inter-particle distance and the new coordinates. The relative position between particles $i$ and $j$ is
\begin{equation}
\left( \begin{array}{c}
x_j - x_i \\
y_j - y_i \\
z_j - z_i
\end{array} \right) = {\bf G} \left( \begin{array}{ccc}
R_1 & 0 & 0 \\
0 & R_2 & 0 \\
0 & 0 & R_3
\end{array} \right)  {\bf G'}^{\rm T} \left( \begin{array}{c}
a_j - a_i \\
b_j - b_i \\
c_j - c_i
\end{array} \right)\, .
\end{equation}
The square of this vector is not a function of the first rotation $\bf G$, but just of the scale matrix and the second rotation matrix
\begin{equation}
r_{ij}^2 = (a_j - a_i \ b_j - b_i \ c_j - c_i) {\bf A} \left( \begin{array}{c}
a_j - a_i \\
b_j - b_i \\
c_j - c_i
\end{array} \right)\, .
\end{equation}

where $\bf A$ is the symmetric matrix
\begin{equation}
{\bf A}  = \left( \begin{array}{ccc}
A_{11} & A_{12} & A_{13} \\
A_{12} & A_{22} & A_{23} \\
A_{13} & A_{23} & A_{33}
\end{array} \right) = {\bf G'} \left( \begin{array}{ccc}
R_1^2 & 0 & 0 \\
0 & R_2^2 & 0 \\
0 & 0 & R_3^2
\end{array} \right) {\bf G'}^{\rm T} \, .
\end{equation}
The six distances are thus functions of six components of matrix $A$ or equivalently, are functions of the six independent coordinates in the scales $R_i$, and the rotation $\bf G'$.

We also compute the kinetic energy as a function of the new coordinates, which is given by
$$
T = {\mu \over 2} \left[ \sum_{i=1}^3 \dot{R_i}^2 - 4( R_2 R_3 \omega_1 \Omega_1 + R_3 R_1 \omega_2 \Omega_2 + R_1 R_2 \omega_3 \Omega_3) \right. +
$$
$$
\omega^{\rm T} \left(\begin{array}{ccc}
R_2^2 + R_3^2 & 0 & 0 \\
0 & R_3^2 + R_1^2 & 0 \\
0 & 0 & R_1^2 + R_2^2
\end{array} \right) \omega +
$$
\begin{equation}
\left.\Omega^{\rm T} \left(\begin{array}{ccc}
R_2^2 + R_3^2 & 0 & 0 \\
0 & R_3^2 + R_1^2 & 0 \\
0 & 0 & R_1^2 + R_2^2
\end{array} \right) \Omega  \right]\, ,
\end{equation}
where $\omega = (\omega_1, \omega_2, \omega_3)$ is the angular velocity vector of the first rotation $\bf G$, and $\Omega = (\Omega_1, \Omega_2, \Omega_3)$ is the corresponding angular velocity vector of the second rotation $\bf G'$.

\section{Equations of Motion}
The equations of motion follow from the Lagrange equations derived from the Lagrangian $T - V$ as presented in any standard text on Mechanics \cite{ll}, \cite{js}.

However, the three coordinates related to the first rotation produces Lagrange equations that imply, when the potential energy is a function only of the distances, conservation of the angular momentum vector in the inertial system
\begin{equation}
{\bf G} \; \, {\bf L}\, ,
\end{equation}
where $\bf L$ is the angular momentum in the principal moments of inertia frame
$$
{\bf L} = \frac{\partial T}{\partial \omega}
$$
\begin{equation}
= \mu \left(\begin{array}{c}
(R_2^2 + R_3^2) \omega_1 \\
(R_3^2 + R_1^2) \omega_2 \\
(R_1^2 + R_2^2) \omega_3
\end{array} \right) - 2 \mu \left(\begin{array}{c}
R_2 R_3 \Omega_1 \\
R_3 R_1 \Omega_2 \\
R_1 R_2 \Omega_3
\end{array} \right)\, .
\end{equation}

This conservation leads to three first order equations forming, for this four-body problem a generalization of the Euler equations (valid for the rotation of a rigid body), namely
$$
\frac{d}{d t} \left( \begin{array}{c}
\mu (R_2^2 + R_3^2) \omega_1 - 2 \mu R_2 R_3 \Omega_1 \\
\mu (R_3^2 + R_1^2) \omega_2 - 2 \mu R_3 R_1 \Omega_2 \\
\mu (R_1^2 + R_2^2) \omega_3 - 2 \mu R_1 R_2 \Omega_3
\end{array} \right) =
$$
\begin{equation}
\left( \begin{array}{ccc}
\mu (R_3^2 - R_1^2) \omega_2 \omega_3 + 2 \mu R_1 (R_2 \omega_2 \Omega_3 - R_3 \omega_3 \Omega_2) \\
\mu (R_1^2 - R_2^2) \omega_3 \omega_1 + 2 \mu R_2 (R_3 \omega_3 \Omega_1 - R_1 \omega_1 \Omega_3) \\
\mu (R_2^2 - R_3^2) \omega_1 \omega_2 + 2 \mu R_3 (R_1 \omega_1 \Omega_2 - R_2 \omega_2 \Omega_1)
\end{array}\right)\, . \label{euler}
\end{equation}

The so called \textit{elimination of the nodes} in the Three-Body problem \cite{wh}, has a similar representation in this coordinates for the Four-Body problem by means of the equation that equals the angular momentum vector in the principal moments of inertia frame to the rotation of a constant vector, which may be written in terms of two Euler angles
\begin{equation}
\mu \left(\begin{array}{c}
(R_2^2 + R_3^2) \omega_1 \\
(R_3^2 + R_1^2) \omega_2 \\
(R_1^2 + R_2^2) \omega_3
\end{array} \right) - 2 \mu \left(\begin{array}{c}
R_2 R_3 \Omega_1 \\
R_3 R_1 \Omega_2 \\
R_1 R_2 \Omega_3
\end{array} \right) = \ell {\bf G}^{\rm T} \left( \begin{array}{c}
0 \\
0 \\
1
\end{array} \right)\, ,
\end{equation}
where $\ell$ is the magnitude of the conserved angular momentum.

The Lagrangian equations of motion for the three scale coordinates are
\begin{equation}
\mu \frac{d^2}{d t^2} R_1 + 2 \mu [R_2 \omega_3 \Omega_3 + R_3 \omega_2 \Omega_2 ] + \mu R_1 (\omega_2^2 + \omega_3^2 +\Omega_2^2 + \Omega_3^2) = - \frac{\partial V}{\partial R_1} \, ,
\end{equation}
\begin{equation}
\mu \frac{d^2}{d t^2} R_2 + 2 \mu [R_3 \omega_1 \Omega_1 + R_1 \omega_3 \Omega_3 ] + \mu R_2 (\omega_3^2 + \omega_1^2 +\Omega_3^2 + \Omega_1^2) = - \frac{\partial V}{\partial R_2} \, ,
\end{equation}
and
\begin{equation}
\mu \frac{d^2}{d t^2} R_3 + 2 \mu [R_1 \omega_2 \Omega_2 + R_2 \omega_1 \Omega_1 ] + \mu R_3 (\omega_1^2 + \omega_2^2 + \Omega_1^2 + \Omega_2^2) = - \frac{\partial V}{\partial R_3} \, .
\end{equation}

The three equations of motion for the three coordinates associated with the second rotation $\bf G'$ are written as an Euler equation similar to the one found for the first rotation, although the internal angular momentum is not conserved because of the presence of an internal torque
$$
\frac{d}{d t} \left( \begin{array}{c}
\mu (R_2^2 + R_3^2) \Omega_1 - 2 \mu R_2 R_3 \omega_1 \\
\mu (R_3^2 + R_1^2) \Omega_2 - 2 \mu R_3 R_1 \omega_2 \\
\mu (R_1^2 + R_2^2) \Omega_3 - 2 \mu R_1 R_2 \omega_3
\end{array} \right) = \left( \begin{array}{c}
K_1 \\
K_2 \\
K_3
\end{array} \right) +
$$
\begin{equation}
\left( \begin{array}{ccc}
\mu (R_3^2 - R_1^2) \Omega_2 \Omega_3 - 2 \mu R_1 (R_2 \omega_2 \Omega_3 - R_3 \omega_3 \Omega_2) \\
\mu (R_1^2 - R_2^2) \Omega_3 \Omega_1 - 2 \mu R_2 (R_3 \omega_3 \Omega_1 - R_1 \omega_1 \Omega_3) \\
\mu (R_2^2 - R_3^2) \Omega_1 \Omega_2 - 2 \mu R_3 (R_1 \omega_1 \Omega_2 - R_2 \omega_2 \Omega_1)
\end{array}\right)\, ,
\end{equation}
where $K_1, K_2, K_3$ are the components of the internal torque $\bf K$ which is expressed in terms of the derivatives of the potential energy with respect to the three independent coordinates $q_j$ in the rotation $\bf G'$ and the three vectors ${\bf c}_j$ that appear in the expression of the angular velocity in terms of the same coordinates
\begin{equation}
\Omega = \sum_{j=1}^3 {\bf c}_j \dot{q}_j\, ,
\end{equation}
where the vectors ${\bf c}_j$ are generally functions of the coordinates $q_j$.

The internal torque is determined by the equations
\begin{equation}
{\bf K} \cdot {\bf c}_j = \frac{\partial V}{\partial q_j}\, .
\end{equation}

There is one  more constant of motion, namely the total energy
$$
E = T + K = V + {\mu \over 2} \left[ \sum_{i=1}^3 \dot{R_i}^2 - 4 ( R_2 R_3 \omega_1 \Omega_1 + R_3 R_1 \omega_2 \Omega_2 + R_1 R_2 \omega_3 \Omega_3) \right. +
$$
$$
\omega^{\rm T} \left(\begin{array}{ccc}
R_2^2 + R_3^2 & 0 & 0 \\
0 & R_3^2 + R_1^2 & 0 \\
0 & 0 & R_1^2 + R_2^2
\end{array} \right) \omega +
$$
\begin{equation}
\left.\Omega^{\rm T} \left(\begin{array}{ccc}
R_2^2 + R_3^2 & 0 & 0 \\
0 & R_3^2 + R_1^2 & 0 \\
0 & 0 & R_1^2 + R_2^2
\end{array} \right) \Omega  \right]\, .
\end{equation}

 \section{The plane problem}
The case with the four particles in a constant plane is an important and old subject \cite{dz}. Our coordinates are now adapted to that case. The third components of the cartesian coordinates of the four particles are zero. The modification of our coordinates (4) for this case is given by two changes: the first one is a rotation by just one angle in the plane of motion; and the second, the scale associated with the third coordinate is set equal to zero, namely
$$
\left(\begin{array}{cccc}
x_1 & x_2 & x_3 & x_4 \\
y_1 & y_2 & y_3 & y_4 \\
0 & 0 & 0 & 0
\end{array} \right) =
$$
\begin{equation}
\left(\begin{array}{ccc}
\cos \psi & -\sin \psi & 0\\
\sin \psi & \cos \psi & 0 \\
0 & 0 & 1
\end{array} \right)
 \left(\begin{array}{ccc}
R_1 & 0 & 0 \\
0 & R_2 & 0 \\
0 & 0 & 0
\end{array} \right) {\bf G'}^{\rm T}
\left(\begin{array}{cccc}
a_1 & a_2 & a_3 & a_4 \\
b_1 & b_2 & b_3 & b_4 \\
c_1 & c_2 & c_3 & c_4
\end{array} \right)\, .
\end{equation}
This equation simplifies to
$$
\left(\begin{array}{cccc}
x_1 & x_2 & x_3 & x_4 \\
y_1 & y_2 & y_3 & y_4
\end{array} \right) =
$$
\begin{equation}
\left(\begin{array}{cc}
\cos \psi & -\sin \psi \\
\sin \psi & \cos \psi
\end{array} \right) \left(\begin{array}{ccc}
R_1 & 0 & 0 \\
0 & R_2 & 0
\end{array} \right) {\bf G'}^{\rm T}
\left(\begin{array}{cccc}
a_1 & a_2 & a_3 & a_4 \\
b_1 & b_2 & b_3 & b_4 \\
c_1 & c_2 & c_3 & c_4
\end{array} \right)\, , \label{flat}
\end{equation}
in terms of six degrees of freedom.

We need three independent coordinates (for example three Euler angles) in $\bf G'$ for the two independent vectors in four dimensions expressed in the base of the three constant vectors $\bf a$, $\bf b$, and $\bf c$, orthogonal to the mass vector.

In order to formulate in a mathematical language the conditions for a plane solution we use the equations introduced by Dziobek \cite{dz}, that have been promoted by many years by A. Albouy and coworkers (see \cite{al} and references therein,) which consists in using the four directed areas of the triangles formed by the particles.

The four (double) directed areas are written in terms of the cartesian coordinates as
$$
S_1 = \left| \begin{array}{ccc}
1 & 1 & 1 \\
x_2 & x_3 & x_4 \\
y_2 & y_3 & y_4
\end{array} \right| \, , \quad S_2 = \left| \begin{array}{ccc}
1 & 1 & 1 \\
x_1 & x_4 & x_3 \\
y_1 & y_4 & y_3
\end{array} \right| \, ,
$$
\begin{equation}
S_3 = \left| \begin{array}{ccc}
1 & 1 & 1 \\
x_1 & x_2 & x_4 \\
y_1 & y_2 & y_4
\end{array} \right| \, , \quad S_4 = \left| \begin{array}{ccc}
1 & 1 & 1 \\
x_1 & x_3 & x_2 \\
y_1 & y_3 & y_2
\end{array} \right| \, ,
\end{equation}
which are the four signed 3 $\times$ 3 minors formed from the matrix
\begin{equation}
\left(\begin{array}{cccc}
1 & 1 & 1 & 1 \\
x_1 & x_2 & x_3 & x_4 \\
y_1 & y_2 & y_3 & y_4
\end{array} \right)\, .
\end{equation}
From now on I will refer to these areas simply as the directed areas.

The addition to the previous matrix of a row equal to any of its three rows produces a square matrix with determinant zero, that implies that the necessary and sufficient conditions to have a constant plane tetrahedron are
\begin{equation}
\sum_{i=1}^4 S_i = 0\, ,\label{plan}
\end{equation}
and
\begin{equation}
\sum_{i=1}^4 S_i x_i = 0\, , \quad \sum_{i=1}^4 S_i y_i = 0\, .\label{afin}
\end{equation}
The two last equations are summarized by the zero vector condition
\begin{equation}
\sum_{i=1}^4 S_i {\bf r}_i = {\bf 0}\, .\label{agrupo}
\end{equation}

An expression for the three directed areas in terms of the previous coordinates follows
\begin{equation}
\left( \begin{array}{c}
S_1 \\
S_2 \\
S_3 \\
S_4
\end{array} \right) =  C {\bf M E}^{\rm T} {\bf G'} \left( \begin{array}{c}
0 \\
0 \\
1
\end{array} \right) \, , \label{areas}
\end{equation}
where $C$ is a constant with units of area over mass. Equation (\ref{plan}) is satisfied from this expression of the directed areas, upon using that the four-vectors in matrix $\bf E$ are orthogonal to the mass vector. With the substitution of equations (\ref{flat}) and (\ref{areas}) in equations (\ref{afin}) one obtains an identity, independent of coordinates $R_1$, $R_2$, $\psi$, and of the rotation angles of $\bf G'$.

I denote here $\theta$ and $\phi$ the spherical coordinates determining this vector.
\begin{equation}
{\bf G'} \left( \begin{array}{c}
0 \\
0 \\
1
\end{array} \right) = \left( \begin{array}{c}
\sin \theta \cos \phi \\
\sin \theta \sin \phi \\
\cos \theta
\end{array} \right)\, ,
\end{equation}

Given the four masses, the four directed areas of the four particles are functions of this unit vector direction only, up to a multiplicative constant $C$ depending on the choice of physical units. Other form equivalent to the plane condition (\ref{agrupo}) is also published in reference \cite {pl}.

In the plane case the angular momentum has a constant direction orthogonal to the plane and of magnitude
\begin{equation}
P_\psi = \frac{\partial T}{\partial \dot{\psi}} = \mu [\dot{\psi}(R_1^2 + R_2^2) - 2 R_1 R_2 \Omega_3]\, . \label{mom}
\end{equation}
The kinetic energy becomes
$$
T = {\mu \over 2} \left[ \sum_{i=1}^2 \dot{R_i}^2 - 4( R_1 R_2 \dot{\psi} \Omega_3)+ \dot{\psi}^2 (R_1^2 + R_2^2) \right.+
$$
\begin{equation}
\left.\Omega^{\rm T} \left(\begin{array}{ccc}
R_2^2 & 0 & 0 \\
0 & R_1^2 & 0 \\
0 & 0 & R_1^2 + R_2^2
\end{array} \right) \Omega  \right]\, .
\end{equation}

Now I substitute polar coordinates for the $R_1$ and $R_2$ coordinates
\begin{equation}
R_1 = R \cos \alpha \, , \quad R_2 = R \sin \alpha\, .
\end{equation}
Writing the kinetic energy in terms of the angular momentum constant of motion instead of the $\dot{\psi}$ velocity lead us to
\begin{equation}
T = {\mu \over 2} \left[ \dot{R}^2 + R^2 \left(\dot{\alpha}^2 + \Omega_3^2 \cos^2 (2 \alpha) + \Omega_1^2 \sin^2 \alpha + \Omega_2^2 \cos^2 \alpha \right) \right] +\frac{P_\psi^2}{2 \mu R^2}\, .
\end{equation}
The energy conservation is thus expressed as
\begin{equation}
E = {\mu \over 2} \left[ \dot{R}^2 + R^2 \left(\dot{\alpha}^2 + \Omega_3^2 \cos^2 (2 \alpha) + \Omega_1^2 \sin^2 \alpha + \Omega_2^2 \cos^2 \alpha \right) \right] + \frac{P_\psi^2}{2 \mu R^2} + V \, , \label{ene}
\end{equation}
where $V$ represents the potential energy.

\section{Central configurations}

The non collinear planar central configurations are characterized in our coordinates by constant values of the $\bf G'$ matrix and of the coordinate $\alpha$ associated to the constant value of the ratio $R_1/R_2$. For these cases the angular velocity vector $\Omega$ is the null vector, the angular velocity $\dot{\alpha}$ is also zero and the conservation equations of moment and energy, (\ref{mom}) and (\ref{ene}) respectively, become
\begin{equation}
P_\psi = \mu \dot{\psi} R^2 \, .
\end{equation}
and
\begin{equation}
E = {\mu \over 2} \dot{R}^2 +\frac{P_\psi^2}{2 \mu R^2} + V \, .
\end{equation}
Note that the potential function $V$ becomes equal to a constant divided by coordinate $R$.

These equations are identical to similar equations obtained for the Euler and Lagrange central configurations of the Three-Body problem \cite{bp}. They are formally the same as the equations for the conics in the Two-Body problem in terms of the radius $R$ and the true anomaly $\psi$.

 The constant values of the $\bf G'$ matrix and angle $\alpha$ referred to above are not arbitrary but they are determined by three independent quantities as discussed in what follows.

 I begin with the approach by Dziobek \cite{dz}. The Four-Body central configurations are determined as critical points of the potential energy with a fixed total inertia moment. The planar solutions with zero volume but finite area are obtained taking in account that the variational equation is modified adding the restriction of planar motion. This condition is obtained by Dziobek \cite{dz} from the derivative of the Cayley-Menger determinant with respect to $r_{ij}^2$ that he found to be proportional to the product of the directed areas $S_i S_j$.

It follows that the solution is given in terms of Lagrange's multipliers $\lambda$ and $\sigma$
\begin{equation}
r_{jk}^{-3} = \sigma + \lambda A_j A_k\, ,
\end{equation}
where $A_j = S_j/m_j$ are weighted areas, quotient of the directed area divided by the corresponding mass. This equation was presented by Dziobek \cite{dz}. A proof was published by Moeckel \cite{mo}, and using a different approach to the same problem, deduced by Albouy \cite{al}. A new proof of the equation was obtained in a different approach by Pi\~na and Lonngi \cite{pl}.

It follows from (\ref{areas}) that in a planar solution the weighted directed areas are expressed as
\begin{equation}
\left( \begin{array}{c}
A_1 \\
A_2 \\
A_3 \\
A_4
\end{array} \right) =  C {\bf E}^{\rm T} {\bf G'} \left( \begin{array}{c}
0 \\
0 \\
1
\end{array} \right) = C {\bf E}^{\rm T} \left( \begin{array}{c}
\sin \theta \cos \phi \\
\sin \theta \sin \phi \\
\cos \theta
\end{array} \right)\, . \label{sec}
\end{equation}
The weighted directed areas are up to a normalization factor equal to the third rotated coordinate of the rigid tetrahedra. In terms of vectors (15-18) and angles $\theta$ and $\phi$ this equation is expressed as
\begin{equation}
Aj = C (a_j \sin \theta \cos \phi + b_j \sin \theta \sin \phi + c_j \cos \theta) \, .
\end{equation}
The explicit expressions choosing $C = \sqrt{(m-m_1)/\mu}$ are
$$
A_1 = \frac{m - m1}{\sqrt{m_1 m}} \cos \theta \, ,
$$
$$
A_2 = - \sqrt{\frac{m_1}{m}} \cos \theta + \sqrt{\frac{m_3 + m_4}{m_2}} \sin \theta \sin \phi \, ,
$$
$$
A_3 = - \sqrt{\frac{m_1}{m}} \cos \theta - \sqrt{\frac{m_2}{m_3 + m_4}} \sin \theta \sin \phi + \sqrt{\frac{m_4 (m - m_1)}{m_3 (m_3 + m_4)}} \sin \theta \cos \phi
$$
\begin{equation}
A_4 = - \sqrt{\frac{m_1}{m}} \cos \theta - \sqrt{\frac{m_2}{m_3 + m_4}} \sin \theta \sin \phi - \sqrt{\frac{m_3 (m - m_1)}{m_4 (m_3 + m_4)}} \sin \theta \cos \phi\, .\label{sfero}
\end{equation}

Since the lengths and masses are defined up to arbitrary units, we assume \cite{pl}, with no loss of generality, that the parameter $\sigma$ equals unity
\begin{equation}
r_{jk}^{-3} = 1 + \lambda A_j A_k\, \quad (j \neq k).
\end{equation}

This implies, according to \cite{pl} that the unit of distance always obeys the restriction
\begin{equation}
\sigma = \frac{\sum\limits_{j>k} m_j m_k / r_{jk}}{\sum\limits_{j>k} m_j m_k r_{jk}^2} = 1\, .
\end{equation}

In the paper by Pi\~na and Lonngi \cite{pl} the directed weighted areas are known as four given constants. The previous equation then gives the distances as functions of the unknown parameter $\lambda$. Through them, the areas of the four triangles become functions of $\lambda$, that should obey the necessary restrictions (\ref{plan}), (\ref{afin}), to verify that one has a planar solution. This restriction allows in many cases to determine the value of $\lambda$ and hence the values of the six distances and the four masses. This is an implicit way to deduce planar central configurations with four masses.

In this paper we assume the four masses are known from the beginning, the four weighted areas are then determined by expressions (\ref{sfero}) in terms of the two tuning variables $\theta$ and $\phi$. Given particular values of these two angles, become determined the four constants $A_j$, until a multiplicative factor, which in turn produces a computed central configuration with given distances and masses. These computed masses are not in general equal (or proportional) to the starting values used to compute the orthocentric tetrahedron. The two angles are then tuned until a numerical coincidence is produced between the given and the computed masses. The distances between particles, computed for this last central configuration, correspond to the given masses.

\begin{figure}
\centering
\scalebox{0.6}{\includegraphics{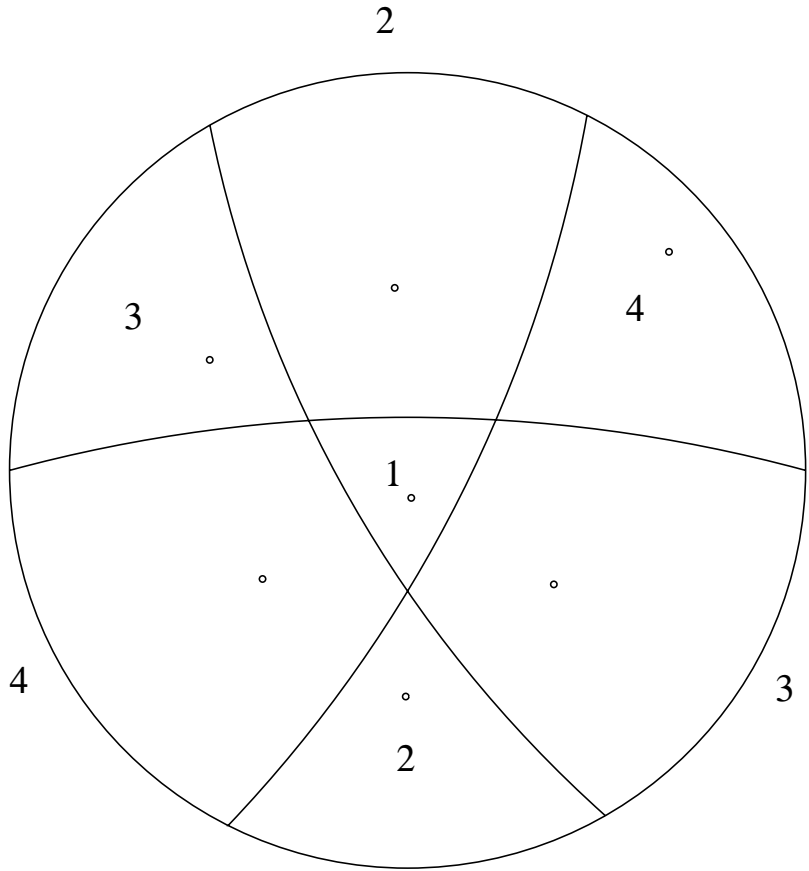}}

\caption{Spherographic projection of the hemisphere of the two angles motion of the orthocentric tetrahedron. The great circles represent the positions where three particles are collinear. The four spherical triangles are concave open sets labeled by the particle at the interior of the triangle. The spherical rectangular open sets correspond to convex configurations with the same order that the neighbor triangles. The isolated points are at the angles where a central configuration has been computed. The values of the masses are $m_1=10, m_2=13, m_3=15, m_4=17$. }
\end{figure}

Choosing the values of the masses as $m_1 = 10, m_2 = 13, m_3 = 15, m_4 = 17$, I have been able to compute seven central configurations.

\noindent A concave with $m_1$ at the interior for $\lambda = -2.32656490060845$
$$
\begin{array}{ll}
\theta = 0.139240050165164 & \phi = 4.8453912490189 \\
r_{31} = 0.749352173668766 & r_{42} = 1.28360719430431 \\
r_{41} = 0.730065912777101 & r_{23} = 1.22971324110202 \\
r_{12} = 0.639643905964532 & r_{43} = 1.10420199339901
\end{array}
$$
A concave with $m_2$ at the interior for $\lambda = -1.82536262997443$
$$
\begin{array}{ll}
\theta = 1.03401714345126 & \phi = 4.70428233126543 \\
r_{23} = 0.804939324544205 & r_{41} = 1.36218205283364 \\
r_{42} = 0.798422993763499 & r_{31} = 1.32975926553097 \\
r_{12} = 0.640524087661698 & r_{43} = 1.07615875773082
\end{array}
$$
A concave with $m_3$ at the interior for $\lambda = -1.6355867896085$
$$
\begin{array}{ll}
\theta = 1.00916705438091 & \phi = 2.69053445366272 \\
r_{23} = 0.819159125672848 & r_{41} = 1.41307775730063 \\
r_{43} = 0.807449041118773 & r_{12} = 1.34385908526924 \\
r_{31} = 0.666344917442357 & r_{42} = 1.08738922587974
\end{array}
$$
A concave with $m_4$ at the interior for $\lambda = -1.55004225977105$
$$
\begin{array}{ll}
\theta = 1.33726059387504 & \phi = 0.589297944199426 \\
r_{41} = 0.80384936609868 & r_{23} = 1.44896029979243 \\
r_{43} = 0.768450306675012 & r_{12} = 1.27362211521548 \\
r_{42} = 0.730824120911527 & r_{31} = 1.18407793737811
\end{array}
$$
A convex with $r_{41}$ at the diagonal for $\lambda = -0.59067068058041$
$$
\begin{array}{ll}
\theta = 0.833903746254753 & \phi = 3.74619198490858 \\
r_{41} = 1.26246852646001 & r_{23} = 1.21353489669217 \\
r_{12} = 0.834421340108319 & r_{43} = 0.914579880279118 \\
r_{31} = 0.855040198605932 & r_{42} = 0.900505139131835
\end{array}
$$
A convex with $r_{42}$ at the diagonal for $\lambda = -0.571602583650311$
$$
\begin{array}{ll}
\theta = 0.849784880696665 & \phi = 5.66013081680728 \\
r_{42} = 1.1712284103105 & r_{31} = 1.31084659340392 \\
r_{12} = 0.837603049326044 & r_{43} = 0.916446768964837 \\
r_{41} = 0.874842967906726 & r_{23} = 0.888555880648135
\end{array}
$$
A convex with $r_{43}$ at the diagonal for $\lambda = -0.532931485997706$
$$
\begin{array}{ll}
\theta = 0.861931053448714 & \phi = 1.64118840218233 \\
r_{43} = 1.13115400784062 & r_{12} = 1.36414262641605 \\
r_{31} = 0.863912312413253 & r_{42} = 0.907050181187521 \\
r_{41} = 0.880188793100079 & r_{23} = 0.893570787310162
\end{array}
$$

\section{The case with two equal masses}

When two masses are equal, the other two generally different, the orthocentric tetrahedron become symmetric with respect to a symmetry plane containing the two (possible) different masses and the center of mass. When the tetrahedron is projected orthogonal to a direction contained in the symmetry plane the configuration becomes with the kite symmetry. Central configurations with two equal masses and kite symmetry are allowed if and only if the plane of symmetry is projected in a line. Such particular case is considered rotating the symmetry plane by a single tuning angle until the given masses coincide with the computed ones. It is only relevant half a circle of this angle. Three cases are possible separated in the semicircle by three contiguous different open sectors. Two concave sectors and one convex sector.

One illustrate this case computing central configurations for the masses
$m_1=8, m_2=10, m_3=m_4=9$. Two different symmetric concave central configurations were found with the largest mass in the interior for $\lambda = - 1.938380387574535$
$$
\begin{array}{ll}
\theta = 0.053846821928071 & \phi = \pi/2 \\
r_{12} = 0.771017850970647 & r_{43} = 1.37174365887618 \\
r_{31} = 0.737687261978754 & = r_{41} \\
r_{23} = 1.24798073364597 & = r_{42}
\end{array}
$$
and for $\lambda = -1.9051833373390781$
$$
\begin{array}{ll}
\theta = 0.131677055692469 & \phi = 3 \pi/2 \\
r_{12} = 0.704334104242389 & r_{43} = 1.16065690193433 \\
r_{31} = 0.777736225967895 & = r_{41}  \\
r_{23} = 1.35289892174911 & = r_{42}
\end{array}
$$
and one convex for $\lambda = -0.6629645165685381$$$
\begin{array}{ll}
\theta = 1.00225261834462 & \phi = \pi/2 \\
r_{43} = 1.2388066072406 & r_{12} = 1.24313752444539 \\
r_{31} = 0.891030948126205 & = r_{41}  \\
r_{23} = 0.864175379988576 & = r_{42}.
\end{array}
$$

No concave solution was found for the lowest mass $m_1 = 8$ in the interior.

Non symmetric central configurations do exist for the case of two equal masses. For the given masses $m_1=8, m_2=10, m_3=m_4=9$ one has the non symmetric convex central configuration obtained by searching with two angles as tuning variables with $r_{23}$ at the diagonal, and the two equal masses not on a single diagonal, for $\lambda = -0.65459120484497$
$$
\begin{array}{ll}
\theta = 0.991809150592805 & \phi = 3.71189655731751 \\
r_{23} = 1.28383284893302 & r_{41} = 1.20043491498806 \\
r_{42} = 0.865256456835286 & r_{31} = 0.891936018662304 \\
r_{12} = 0.878638767853787 & r_{43} = 0.879635445676687
\end{array}
$$
and the the non symmetric concave central configuration with one of the equal masses at the interior for $\lambda = -2.0782713344769$
$$
\begin{array}{ll}
\theta = 1.2731702593358 & \phi = 2.4920236726443 \\
r_{23} = 0.681773343374296 & r_{41} = 1.13338465823949 \\
r_{31} = 0.767118374257656 & r_{42} = 1.31028169747428 \\
r_{43} = 0.776744110452106 & r_{12} = 1.35194750927076
\end{array}
$$

\end{document}